\documentclass[conference]{IEEEtran}
\IEEEoverridecommandlockouts
\usepackage{cite}
\usepackage{amsmath,amssymb,amsfonts}
\usepackage{algorithmic}
\usepackage{graphicx}
\usepackage{textcomp}
\usepackage{xcolor}
\usepackage{multicol}

\usepackage{amsmath,amssymb,amsfonts}
\usepackage{algorithmic,algorithm}

\usepackage{lipsum,graphicx,subcaption}
\def\BibTeX{{\rm B\kern-.05em{\sc i\kern-.025em b}\kern-.08em
    T\kern-.1667em\lower.7ex\hbox{E}\kern-.125emX}}
    
\begin{document}
\nocite{*}
\title{A Fuzzy Logic Controller for Tasks Scheduling Using Unreliable Cloud Resources\\
}

\author{\IEEEauthorblockN{Panagiotis Oikonomou}
\IEEEauthorblockA{\textit{Computer Science and Engineering} \\
\textit{Southern Univ. of Science and Technology}\\
Shenzhen, China \\
oikonomoup2019@mail.sustech.edu.cn}
\and
\IEEEauthorblockN{Kostas Kolomvatsos}
\IEEEauthorblockA{\textit{Computer Science and Telecomm.} \\
\textit{University of Thessaly}\\
Lamia, Greece \\
kostasks@uth.gr}
\and
\IEEEauthorblockN{Nikos Tziritas}
\IEEEauthorblockA{\textit{Computer Science and Telecomm.} \\
\textit{University of Thessaly}\\
Lamia, Greece \\
nitzirit@uth.gr}
\and
\IEEEauthorblockN{Georgios Theodoropoulos}
\IEEEauthorblockA{\textit{Computer Science and Engineering} \\
\textit{Southern Univ. of Science and Technology}\\
Shenzhen, China \\
georgios@sustech.edu.cn}
\and
\IEEEauthorblockN{Thanasis Loukopoulos}
\IEEEauthorblockA{\textit{Comp. Science and Biomedical Informatics} \\
\textit{University of Thessaly}\\
Lamia, Greece \\
luke@uth.gr}
\and
\IEEEauthorblockN{Georgios Stamoulis}
\IEEEauthorblockA{\textit{Electrical and Computer Engineering} \\
\textit{University of Thessaly}\\
Volos, Greece \\
georges@uth.gr}

}

\maketitle

\begin{abstract}
The Cloud infrastructure offers to end users a broad set of heterogenous computational resources using the pay-as-you-go model. These virtualized resources can be provisioned using different pricing models like the unreliable model where resources are provided at a fraction of the cost but with no guarantee for an uninterrupted processing. However, the enormous gamut of opportunities comes with a great caveat as resource management and scheduling decisions are increasingly complicated. Moreover, the presented uncertainty in optimally selecting resources has also a negatively impact on the quality of solutions delivered by scheduling algorithms. In this paper, we present a dynamic scheduling algorithm (i.e., the Uncertainty-Driven Scheduling - UDS algorithm) for the management of scientific workflows in Cloud. Our model minimizes both the makespan and the monetary cost by dynamically selecting reliable or unreliable virtualized resources. For covering the uncertainty in decision making, we adopt a Fuzzy Logic Controller (FLC) to derive the pricing model of the resources that will host every task. We evaluate the performance of the proposed algorithm using real workflow applications being tested under the assumption of different probabilities regarding the revocation of unreliable resources. Numerical results depict the performance of the proposed approach and a comparative assessment reveals the position of the paper in the relevant literature.
\end{abstract}


\section{Introduction}
Workflow scheduling is the process of mapping inter-connected tasks on heterogeneous resources (resources with different computational and storage capabilities). This is a fundamental and well-studied problem in computing environments such as Grid and Cluster Computing \cite{wu2015workflow}. Research centers (e.g., NASA, earthquake-epigenomic centers) are utilizing such computing environments to distribute the workload of complex and heavy load scientific experiments. The last decade, there is a growing interest on scheduling algorithms applied for such workflows in the Cloud \cite{rodriguez2017taxonomy}. The substantial amount of resources, the variety of CPU platforms (vCPUs) as well as the zero cost for management/maintenance has made Cloud the most suitable environment for the execution of computation intensive applications. However, challenges like the pay-as-you go model and data-transfer costs can be a obstacle to Cloud’s potentials \cite{wu2015workflow}. 
A main difference between Cloud and Cluster Computing is that users have to pay for the duration that resources are utilized. In addition, in Cloud environments the performance of resources varies. The above is caused by the resource sharing between Virtual Machines (VMs) hosted in the same physical machine.

Amazon Web Services (AWS) is a typical example of a Cloud provider that offers multiple services. 
The main categories of services are (i) on-demand and (ii) reserved resources instances. On-demand instances have a fixed price for each hour of use while reserved instances have a cheaper per-hour price than the on-demand services, however, users must lease them for long periods of time (more than 1 year).
Amazon was one of the first providers that announced the disposal of unused capacity with a significant discount (around 80\% compared to on-demand services). This new type of services is referred as a spot instance. 
Large Cloud providers followed the Amazon and offer spare capacity at a discount as well. Google Compute Engine (GCE) and Alibaba launch Preemptible Virtual Machines (VMs) while Azure offers a low-priority VMS. However, from the user’s standpoint, using such virtualized resources comes with a major caveat. Instances may be revoked by the provider at any time as their capacity is needed to execute other (preemptive) applications. For instance, Google sends a preemption notice thirty seconds before termination. Usually, preemptive instances are terminated after 24 hours of use. 
Spot services can be acquired via a biding policy through an auction-like market and greedily, the user with the maximum bid acquires the instance. We have to notice that spot instances are revoked if user’s bid price is lower than market price. 

A typical scenario in the Cloud environment is that a user wants to execute a workflow application in the minimum time and cost. Since these requirements are orthogonal in nature, users are confronted with a time versus cost dilemma under the constraints imposed by the workflow and the provider. This paper aims at finding a solution for this challenging dilemma. A straightforward solution to minimize monetary cost is to use exclusively unreliable virtualized resources, i.e., completely rely on spot instances. A key problem to this solution is that the resource availability and the uninterrupted execution of the workflow are not guaranteed. The adopted scheduler should constantly monitor the queues of virtualized resources and backup task’s progress even if the possibility of premature termination is relatively low. 
Minimizing the workflow’s execution time (makespan) is a subject that has been extensively studied by the research community. State-of-the-art algorithms like HEFT, CPOP \cite{topcuoglu2002performance} and DCP  \cite{kwok1996dynamic} are effective to minimize the makespan, however, the monetary cost is disregarded. Mapping tasks exclusively to reliable (on-demand) virtualized resources towards securing the uninterrupted execution of a workflow results in enormous monetary costs compared to a schedule that considers the pay-as-you-go model.

In this paper, we focus on investigating dynamic scheduling approaches like in \cite{oikonomou1}, \cite{oikonomou2} whereby it is decided in which virtualized resource each task should be assigned. The decisions are made based on the current state of the system and the workflow execution requirements. An algorithm that optimizes simultaneously the workflow’s execution time (makespan) and the monetary cost using a mixture of reliable and unreliable resources is proposed. To mitigate the performance variation of Cloud environment as well as the unstable nature of unreliable virtualized resources, we propose a `fast', however, efficient technique that covers the uncertainty present into our scenario. The discussed uncertainty deals with the most appropriate resource that should selected to host every task under the target of minimizing the execution time and and monetary costs. We adopt the principles of Fuzzy  Logic (FL) to handle the uncertainty of the selection process and the definition of an efficient decision making
thresholds. FL is widely adopted in many applications domains (tasks scheduling among them \cite{paper1}) as the appropriate theory/technology for dealing with uncertainty in decision making (e.g., \cite{paper2}, \cite{paper3}, \cite{paper4}). We depart from the relevant literature and 
avoid using `crisp' thresholds in decision making.
For instance, other efforts in the field target to meet 
specific crisp thresholds for deadlines and budget constraints when deciding tasks assignments. 
The intuition behind our approach is two-fold: First, we always seek to minimize the adopted parameters alleviating users from the burden of defining specific thresholds and, secondly, our algorithm takes into consideration multiple parameters (makespan, monetary cost) at the same time leading to a multi-objective decision making.
To the best of our knowledge this is one of the first efforts that deal with the problem of scheduling scientific workflows using multiple unreliable virtualized resources without taking into consideration any Quality of Service (QoS) constraints. The following list reports on the contribution of our work:
 \begin{itemize}
\item We propose a model that captures the heterogeneity of the Cloud environment. Both reliable and unreliable virtualized resources with different processing capabilities can be provisioned exhibiting different interruption probability depending on the popularity of each resource;
\item We provide a FL Controller (FLC) to decide on the type of the resources we have to adopt to host each task of the desired workflow, thus, we manage the uncertainty related to the discussed decision-making problem;
\item We perform an extensive experimental evaluation of the proposed model and simulate the execution of 5 real-world scientific workflows. The configuration of the adopted virtualized resources is based on realistic assumptions (AWS).
\end{itemize}

The rest of the paper is organized as follows: Section \ref{section:relatedWork} discuss the related work. System model and problem formulation are illustrated in Section \ref{section:systemproblem}. The proposed algorithm is presented in Section \ref{section:algorithm}. Concluded remarks are discussed in Section \ref{section:experiments}. Our conclusions are drawn in the final section \ref{section:conclusions}.

\section{Related Work}
\label{section:relatedWork}
A growing body of literature has examined the problem of scheduling workflows in Cluster and Cloud computing. The majority of resource provision techniques are guided by QoS constraints defined by users. There are two basic constraints, i.e., the deadline \cite{martinez2019planning}, \cite{suguna2018heuristic}, \cite{chen2017cloud}, \cite{poola2016enhancing}, \cite{ghafarian2015cloud}, \cite{poola2014fault} and the monetary budget \cite{monge2020cmi}, \cite{tordini2018scientific}, \cite{jung2014workflow}, \cite{monge2014adaptive}. 
While the deadline constraint is usually satisfied, unreliable resources are used to the maximum extent to 
limit the monetary costs. 
In the following paragraphs, we categorize the relevant 
efforts found in the literature and describe the most
representative models. 

\textbf{Heuristic Workflow Scheduling}.
In \cite{poola2014fault}, the authors introduce the concept of the Latest Time On-Demand (LTO), i.e., the latest time in which on-demand instances must be used to guarantee the deadline constraint. If the difference between the LTO and the current time is greater than zero (positive slack), tasks are mapped into spot instances (unreliable resources). This ensures that the deadline constraint is satisfied while cost is minimized. Decisions made upon the slack value inspired other works like \cite{poola2014fault}, \cite{ghafarian2015cloud} and \cite{poola2016enhancing}. A just in time workflow scheduling algorithm with deadlines guarantees and cost minimization is presented in \cite{poola2014fault}. A task that arrives before the LTO is scheduled to a spot instance otherwise it is scheduled to an on-demand resource. Only a single spot instance is considered (the cheapest one). The maximization of resources utilization as well as the number of tasks that fulfil QoS constraints is the subject of \cite{ghafarian2015cloud}. Grid resources are adopted as the default candidate solutions. However, if QoS constraints are not satisfied, tasks, along with their predecessors, are executed on spot instances. 
In \cite{chen2017cloud}, two types of tasks are considered; Preemptive tasks are executed exclusively in spot instances while non-preemptive tasks are executed on reliable instances.  
For each task the scheduler scans the entire list of busy resources to find an idle time-block that gives the earliest start time (insertion policy).
A framework for scheduling scientific workflows in a Hybrid Cloud environment (HC) is presented in \cite{tordini2018scientific}. HC consists of multiple Data Centers (DCs) containing both reliable and unreliable VMs. Processing elements within each DC are homogenous while costs for data movement and dynamic resource provisioning are ignored. An execution manager is responsible to monitor tasks executed in revocable VMs and to reschedule them when necessary. 
An extension of the DCP algorithm \cite{kwok1996dynamic} is presented in \cite{ostermann2012impact}. A Grid infrastructure is extended with unreliable public Cloud resources when Grid’s resources are insufficient/unavailable to process the current execution load.

\textbf{Resource Provisioning Techniques}. 
In \cite{monge2014adaptive}, the authors address the problem of auto-scaling spot resources. A new proposed strategy called Spots Instances Aware Autoscaling (SIAA)  
aims at the elimination the makespan and the probability of task failures. According to the available budget, SIAA generates a scaling plan compromised by both reliable and unreliable instances. 
Critical tasks are prioritized first (tasks with small slack time), then, every one of them is scheduled based on the Earliest Finish Time (EFT) policy. 
In \cite{zhou2015monetary}, the authors present a framework for scheduling multiple workflows that offers probabilistic deadline guarantees and monetary cost minimization. 
At runtime, a combination of on-demand and spot instances is generated for every task. 
If the execution of a task on spot instances fails or deadline is not met, on-demand instances are adopted on the fly. 
In \cite{chard2015cost}, the authors present an elastic resource provisioner for the allocation of on-demand and spot instances to workflow tasks. High spot prices defined by users trigger the switch from unreliable to reliable resources. 
In \cite{monge2020cmi}, we can find a discussion on the problem of auto-scaling public resources using a Multi-objective Genetic Algorithm (GA). The makespan, the monetary cost and Out-of-Bid (OOB) errors are considered as the targets of the minimization process. The overall impact of OOB errors is measured using a probabilistic model which takes as inputs the probability of OOB errors occurrence multiplied by the number of the available vCPUs. 

\textbf{Fault-Tolerant Models}. Checkpointing, a mechanism to maintain the reliability of unreliable resources, is introduced in \cite{yi2010reducing}. A load balancing model combined with a GA could decide on the optimal number of tasks within an instance \cite{jung2014workflow}. Fault-tolerance is enforced using a two-threshold (price and time) mechanism. However, only identical sized tasks are considered and Cloud resources are limited to spot instances. In \cite{poola2016enhancing}, the provisioning of spot instances is associated to task duplication, i.e., tasks are marked for duplication when the scheduler detects idle slots that can execute a task replicas. 
If no suitable idle slot exists, a new spot instances is initiated to host replicas. 
In \cite{xu2016multi}, the authors propose a multi-objective GA that minimizes both the makespan and the monetary cost. However, it is assumed that all faults in task’s execution are revocable. Tasks can continue their execution in the allocated spot instance after a while (fault-recovery).

\section{System Model and Problem Formulation}
\label{section:systemproblem}
\subsection{System Model}
We consider the scenario where a user wants to execute a set of dependent tasks (workflow application) in an Infrastructure as a Service (IaaS) Cloud environment. 

\textbf{Definition.}  A workflow is a set of dependent tasks that solve a scientific problem.

Resources are general provisioned as VMs and a VMs pool configuration is defined prior to workflow execution offered by the provider. The cost of leasing such virtualized resources is bounded between the start time of the first task assigned to it and the completion time of the last task assigned to it. The cost is rounded up to the nearest billing cycle. 
A workflow application can be modeled as a Directed Acyclic Graph (DAG) $G=(T,E)$, where $T$ is the set of tasks and $E$ is the set of edges. Each edge $e_{ij}$ represents data dependencies between tasks $t_i$ and $t_j$. Task $t_j$ receives $e_{ij}$ amount of data from its predecessor, i.e., task $t_i$. 
For starting the execution of a task ($t_i)$ the following two conditions must hold true: a) all predecessor ($pred(t_i)$) tasks must finish execution and b) all data from $pred(t_i)$ must be received. $t_i$ is characterized by a processing demand parameter $g_i$ denoting the number of instructions (i.e., MIPS) that must be executed for its completion. A task is called an entry task ($t_{entry}$) when ($pred(t_i) = \emptyset$). Similar, a task without any successor task ($succ(t_i) = \emptyset$) is called an exit task ($t_{exit}$). If more than one entry tasks exist, then, a pseudo task $t_{pseudo}$ is inserted to $G$ as the predecessor task of every entry task. No data is transmitted from $t_{pseudo}$ to any other task. Multiple exit tasks are handled in an analogous manner.

Let $V$ be the set of heterogeneous resources (VMs) forming the aforementioned VM pool configuration. Any virtualized resource can be leased as a reliable or an unreliable instance. Additionally, every VM is associated with a processing capability and incurs in a different cost per use. Let $v_i$ denote the $i$th resource, $r_i$ be the processing power of $v_i$ and $u_i$ be the leasing cost of $v_i$. Precisely, $r_i$ denotes the number of MIPS instructions that can be executed per time unit (i.e., one second) by $v_i$. This also incorporates memory speed, disk size and so on and so forth. Each virtualized resource is also associated with a preemption/interruption probability $p_i$. 

\textbf{Definition.} The preemption/interruption probability $p_{i}$ is the probability of `loosing' the selected virtualized resource leading to the failure of the corresponding task and the need of a re-execution. 

Since reliable resources (on-demand) are irrevocable, $p_i$ is set to zero. Regarding unreliable resources, $p_i$ is a positive number set to unity when the active duration of $v_i$ exceeds one hour (as dictated by GCE). We assume that the tasks that consist of a Workflow are non-preemptive and atomic (must be executed again if they fail). The execution time required to complete the task $t_i$ on resource $v_j$ is calculated by Eq. (\ref{exTime}).
\begin{equation}
\label{exTime}
w_{i,j}=\frac{g_i}{r_j}
\end{equation}
We consider that resources are allocated in the same DC, thus, transferring inbound data is free. DC is assumed to have enough resources to schedule $G$’s tasks. A shared global storage system is considered as a data repository (Amazon S3). Tasks save their outputs and receive their inputs from the same storage system. Since the global storage system is allocated within DCs premises, we consider that data transfer rate between two VMs is constant. Let $T$ be that data transfer rate and $b_{ij}$ be a binary variable, i.e., $b_{ij}=0$ iff  $i=j$ otherwise $b_{ij}=1$. The temporal cost to send data form $t_i$ to $t_j$ ($t_i$ is assigned in $v_i$ while $t_j$ in $v_j$) is expressed by Eq. (\ref{dataCost}).
\begin{equation}
\label{dataCost}
d_{i,j}=\frac{e_{ij}b_{ij}}{r_j}
\end{equation}
$EST_{t_i}^{r_j}$ and $EFT_{t_i}^{r_j}$ are the earliest execution start time and the earliest execution finish time of $t_i$ on $r_j$. The earliest start time of the entry task is zero. For every other task in $G$, the earliest start time and the earliest finish time are calculated recursively as shown in Eq. (\ref{EST}) and Eq. (\ref{EFT}):
\begin{equation}
\label{EST}
EST_{t_i}^{r_j}=max_{{t_z}\in pred(t_i)}\{AFT_{t_z}+d_{zi}\}
\end{equation}
\begin{equation}
\label{EFT}
EFT_{t_i}^{r_j}=w_{ij}+EFT_{t_i}^{r_j}
\end{equation}
However, due to Cloud uncertainties like multi-tenant resource sharing, resource revocation and provision-deprovision delays, EST and EFT can be underestimated. Therefore, we introduce $AST_{t_i}$ and $AFT_{t_i}$ that denote the actual execution start time and the actual finish time of $t_i$. Then, the total elapsed time required to execute $G$ (makespan) is expressed by Eq.(\ref{makespan}): 
\begin{equation}
\label{makespan}
makespan(G)=AFT_{t_{exit}}
\end{equation}
Let $N$ denote the total number of tasks scheduled by $r_i$ and $t_k$ be the $k$th task assuming a total ordering of them $1 \leq k \leq N$. 
Then, the overall execution cost incurred by $r_i$ is calculated by (\ref{overallExCost}), where $\gamma$ is the length of the billing cycle.
\begin{equation}
\label{overallExCost}
c_i=u_i
 \left[  \frac{ AFT_{t_N}-AST_{t_1} } { \gamma }  \right]
\end{equation}

\subsection{Problem Formulation}
Let $X$ be an $|V|×|T|$ binary matrix used to encode task-resources assignments as follows: $X_{ij}=1$ iff $t_i$ is assigned for processing at $v_j$, otherwise $X_{ij}=0$. In our model, 
time is represented with the introduction of $S$ equally sized time slots. 
Let $S_\tau$ be the $\tau$th such time slot, with a corresponding assignment matrix $X_\tau$. 

\textbf{\textit{Problem}}: Find all values in the $X$ total matrices $X_\tau$, so that the objective function $f$ given by Eq.(\ref{objective}) is minimized:  
\begin{equation}
\label{objective}
f=(AFT_{t_{exit}}, \sum_{i=1}^{|V|}c_i) 
\end{equation}
subject to:
\begin{equation} \tag{c1}
\label{c1}
\sum_{j=1}^{|T|}X_{kj}^\tau \leq 1, \forall k,\tau
\end{equation}
\begin{equation}
\label{c2} \tag{c2}
\sum_{i=1}^{|V|}X_{ik}^\tau \leq 1, \forall k,\tau
\end{equation}

\textbf{\textit{Research Challenge:}} 
“\textit{minimize the makespan and the overall monetary cost, i.e., minimize the objective function $f$ provided by Eq.(\ref{objective}) w.r.t. 
the following constraints: (i) a resource cannot execute concurrently more than one tasks Eq.(\ref{c1}), and (ii) a task cannot be assigned to more than one resources Eq.(\ref{c2})} ".

\section{The Uncertainty-Driven Scheduling (UDS) Algorithm}
\label{section:algorithm}
At each time step $\tau$, our algorithm tries to accomplish two goals, i.e., (i) the minimization of the makespan and (ii) the minimization of the monetary cost. To achieve both goals, we introduce the concept of effectiveness ($eff^\tau(M_x,C_x)$) which is reinforced to each scheduling decision. $M_x$ and $C_x$ denote the makespan and monetary cost after applying scheduling plan $x$.

\textbf{Definition.} Effectiveness $eff^\tau(M_x,C_x)$ is defined as the ability of an algorithm to deliver the optimal execution of a workflow in a timely manner after applying scheduling plan $x$.

$eff^\tau(M_x,C_x)$ is measured for both goals as the difference between: (a) the expected performance of the algorithm from the current time $t$ to the finish of the schedule (including every future decision), $eff^\tau(M_{ideal},C_{ideal}), {\tau \in [\tau,finish]}$  and (b) an idealistic performance which actually minimizes both metrics to the maximum extreme possible $eff^\tau(M_{ideal},C_{ideal}), {\tau \in [start,finish]}$. Clearly, if, at the end of the schedule ($\tau=finish$), the difference between the two performances is eliminated (it is close to zero) on both metrics, our algorithm's performance is considered as efficient. 

In general, the problem of assigning tasks to heterogeneous resources is NP-hard \cite{fernandez1989allocating}, thus, not a known algorithm is able to generate the optimal solution within polynomial time. For this reason, we assume that the theoretical optimal performance is accomplished using two well-known greedy algorithms namely HEFT \cite{topcuoglu2002performance} and GreedyCost (GS). 
HEFT is applied upon the makespan metric while GC for the cost metric, respectively. For each task, HEFT selects the resource that results in the earliest finish time while the GC relies on the resource that results in the lower cost. 
For both algorithms, the task to resource mapping is produced in advanced (i.e., they perform a static scheduling) and all resources can be used in an uninterrupted mode, thus, they can produce high-quality schedules. This means that both algorithms consider a $p_{i}$ equal to zero.
For our analysis and experimentation, we consider that $M_{lower}=HEFT_{ideal}$ and $C_{lower}=GC_{ideal}$ represent the lower bound of the makespan and the monetary cost respectively.
\begin{algorithm}[h]
 \caption{The UDS Algorithm}
 \label{alg:xyz_Algorithm}
 \textbf{Input:} Workflow's tasks $T$, Pool of resources $R$, $\theta$, $a$, $b$\\
  \textbf{Output:} $M_{final}$, $C_{final}$ 
 \begin{algorithmic}[1]
 \renewcommand{\algorithmicrequire}{\textbf{Input:}}
  \STATE Calculate $M_{lower}$, $C_{lower}$
  \STATE Call function $eff^t(M_{lower},C_{lower})$
  \STATE $M_{upper}=M_{lower}+a\times M_{lower}$
  \STATE $C_{upper}=C_{lower}+b\times C_{lower}$
  \STATE $Q \leftarrow t_{entry}$
  \WHILE {$Q \neq \emptyset$}
  \STATE Select $t_i$ from $Q$
  \STATE $W = \{t_i\}$, $t_{i}$ is in waiting state
  \STATE Call function $eff^\tau(M_{curr},C_{curr}) \forall t_i \in W$
  \STATE $normM=(M_{curr}^\tau-M_{lower})/(M_{upper}-M_{lower})$
  \STATE $normC=(C_{curr}^\tau-C_{lower})/(C_{upper}-C_{lower})$
  \STATE $PMI_i^\tau=FLC(normM,normC)$
  \IF {($PMI_i^\tau \geq \theta$)}
  \STATE Select a reliable pricing model
  \ELSE
  \STATE Select an unreliable pricing model
  \ENDIF
  \FORALL {$r_j \in R$}
  \STATE Compute $EFT_{t_i}^{r_j}$
  \ENDFOR
  \STATE Schedule $t_i$ to $r_j$ that minimize $EFT_{t_i}$
  \STATE Update $Q$ with $succ(t_i)$ if $\forall t_j \in pred(t_i), AFT_{t_j}+d_{ij} \leq \tau$
  \ENDWHILE
 \end{algorithmic} 
 \end{algorithm}
 
Algorithm \ref{alg:xyz_Algorithm} describes the UDS algorithm. For our scheduling scenario, decisions for each task are made at the runtime i.e., when a task is ready for execution. 
In the resource allocation phase (lines 6-22)
for every ready task ($t_i$), we estimate $eff^\tau(M_{curr},C_{curr})$. To do so, at first, we apply HEFT and GC in a dynamic way (decisions are based on the current time). We should mention that HEFT \& GC consider only tasks that, at $t$, are in waiting state (line 8), i.e., tasks that are not able to run yet because the conditions for running are not in place (precedence constraints). Both HEFT and GC will result in different solutions w.r.t. the makespan and the cost which, in turn, results in different distances from $M_{lower}$  and $C_{lower}$ (line 9). 
Let $M_{curr}^\tau$ and $C_{curr}^\tau$ denote the aforementioned solutions and $M_{upper}$ and $C_{upper}$ be the upper bounds for both $M$ and $C$ as expressed in lines 2 and 3, respectively.  $a$ and $b$ are scalar values. Next, both $M_{curr}^\tau$ and $C_{curr}^\tau$ are normalized in the unity interval based on the aforementioned lower and upper bounds (lines 10, 11).
Clearly, when $C_{curr}^\tau$ is relatively small compared to $C_{curr}^\tau$, then, unreliable VMs should be utilized to reduce the overall monetary cost. The adversary case indicates that reliable VMs should be used. Our goal is to minimize the distance between the solution ($M_{lower}, C_{lower}$) and the one generated by our approach for every ready task ($C_{curr}^\tau, C_{curr}^\tau$). 
However, due to performance fluctuations in the Cloud environment selecting the appropriate virtualized resource (type and computational capabilities) is a challenging task.

\section{The Uncertainty Driven Decision Making}
\label{section:algorithm}

\subsection{The Proposed FLC}
As it is difficult to be aware and define specific thresholds for both metrics (makespan and monetary cost) to support efficient resource allocation and aiming at the management of the ambient uncertainty, we adopt an FLC to support the  final decision related to the selection of the appropriate resources (line 7 of the proposed algorithm). In 
FL systems, the objects of discourse are associated with information which is, or is allowed to be, incomplete, partially true or partially possible. FL deals with incomplete information and provides knowledge representation models, i.e., Fuzzy Set Theory, through which an entity can automatically take decisions during the fulfillment of a task. FL principles express human expert knowledge and enable the automated interpretation of results. The proposed FLC is responsible to handle the uncertainty in decision making and the definition of thresholds for the involved parameters. The FLC is a non-linear mapping between $l$ inputs $u_i\in U_i,i=1,\dots,l$ and $m$ outputs $y_i\in Y_i,i=1\dots,m$. We adopt the Mamdani type of inference \cite{bharathi2008characterization} that utilizes rules as the following: $R_{j}$: IF $u_{1j}$ is $A_{1j}$ AND/OR $u_{2j}$ is $A_{2j}$ AND/OR $\ldots$ AND/OR $u_{lj}$ is $A_{lj}$ THEN $y_{1j}$ is $B_{1j}$ AND $\ldots$ AND $y_{mj}$ is $B_{mj}$, where $R_{j}$ is the $j$th fuzzy rule, $u_{ij} (i = 1, \ldots, l)$ are the inputs of the $j$th rule, $y_{kj} (k = 1, \ldots, m)$ are the outputs and $A_{ij}$, $B_{kj}$ are membership functions usually associated by linguistic terms. 

The proposed FLC has two inputs, i.e., $M_{curr}^\tau$ \& $C_{curr}^\tau$. The single output of the FLC is the Pricing Model Indicator (PMI), $PMI_i^\tau$. When $M_{curr}^\tau\rightarrow1$ (High), it means that there is an increased demand to decrease the makespan, the opposite is true when  $M_{curr}^\tau\rightarrow0$ (Low). When $C_{curr}^\tau\rightarrow1$ (High) then the current scheduling decision suffers from high monetary cost, the opposite stands for $C_{curr}^\tau\rightarrow0$ (Low). Concerning output fuzzy variable $PMI_i^\tau$ a value close to one (High) indicates that reliable resources should be used to decrease the overall execution time of the workflow. On the other hand, a value close to zero (Low) depicts a ‘decrease monetary cost’ decision, thus, task $t_i$ should be executed to unreliable resources. So far, the FLC is capable to decides on the type of resources that must be selected (reliable or unreliable).

For inputs and the output, we consider three linguistic values: Low, Medium, High. A Low value represents that the fuzzy variable takes values close to the lowest limit while a High value depicts the case where the variable takes values close to the upper limit. A Medium value depicts the case where the variable takes values close to the average (e.g., around 0.5). For simplicity, we consider triangular membership functions as they are widely adopted in the literature. However, the proposed framework is generic enough and, thus, one can adopt any membership function that better suits to the application domain.

The proposed FLC receives crisp values for the two inputs, it fuzzifies them and, accordingly, proceeds with the inference process. The inference process involves a set of fuzzy rules that result the best possible value for the output $PMI_i^\tau$. These rules are defined by experts and incorporate the human view on the decision process that we should follow. In Table \ref{rules}, we present the adopted FL rule base. These rules are designed for the specific scenario and exhibit a behavior that resembles human reasoning, e.g., if the monetary cost is high and the execution time is low then allocate current task to an unreliable resource. The final step is the de-fuzzification process in order to derive the final $PMI_i^\tau$ value. When the $PMI_i^\tau$ value is over a pre-defined threshold ($\theta$), task $t_i$ is scheduled for execution to a reliable VM, otherwise is executed to an unreliable VM. Our proposed methodology considers multiple heterogeneous virtualized resources, thus, to conclude on the computational capabilities for the VM that eventually will host $t_i$ we select the one that minimize $t_i$'s finish time the most (lines 18-20). 
\begin{table}[h]

\caption{Fuzzy Logic rule base}
\label{rules}
\centering
\begin{tabular}{l|ll|l}
\hline \hline
No & $M_{curr}^\tau$ & $C_{curr}^\tau$ & $PMI_i^\tau$ \\
\hline
1 & Low          & $\left\lbrace Low, Medium, High \right\rbrace$   & Low           \\
2 & Medium          & Low   & High           \\
3  & Medium           & Medium           & Medium          \\
4  & Medium        & High            & Low           \\
5 & High          & $\left\lbrace Low, Medium, High \right\rbrace$   & High           \\
\hline \hline
\end{tabular}
\end{table}

\section{Experimental Evaluation}
\label{section:experiments}

\subsection{Simulation setup}
\textbf{\textit{Workflow Applications}}. 
We report on the experimental 
evaluation of the proposed model relying on five (5) workflow applications as depicted by \cite{juve2013characterizing}, \cite{bharathi2008characterization}. 
The number of tasks, the execution time of each task as well the amount of data transferred between them is reported in a ‘Directed Acyclic Graph in XML’ (DAX) format. 
Workflows include Montage, LIGO, CyberShake, SIPHT and Epigenomics which are extensively adopted in 
the relevant literature. The discussed workflows `cover' all the basic execution patterns such as pipelining, process, data aggregation, data distribution and data redistribution. Each workflow contains 1,000 tasks.

\textbf{\textit{Virtualized Resources}}.
We consider a Cloud model with a single DC offering VMs of different CPU speeds and prices. 
For each experiment, we consider five (5) reliable and 
five (5) unreliable resources with their  
characteristics being generated upon the Amazon EC2 platform. 
Generic VMs are considered from the US East (Ohio) region in a Linux operating system. 
Table \ref{VMs} presents the adopted VMs characteristics. 
We assume that the execution time of each task provided in the DAX files is on the slowest available VM (a1.medium).
The average bandwidth between the storage system (S3) and VMs is set to 20 Mbps which is the approximate average bandwidth provided by Amazon services \cite{palankar2008amazon}. To measure the performance fluctuations of the adopted VMs, we follow a similar approach as the one presented in \cite{schad2010runtime}, \cite{sahni2015cost}. The performance of VMs varies up to 19\% based on a normal distribution with a mean of 9.5\% and standard deviation of 5\%. The bootup/startup time for each VM (provisioning time) is set to 96.9 seconds \cite{mao2012performance}.

\begin{table}[h]
\caption{VMs characteristics}
\label{VMs}
\begin{tabular}{l|c|r|l|c}
\hline
\hline
Type       & \multicolumn{1}{l|}{vCPUs} & \multicolumn{2}{l|}{\begin{tabular}[c]{@{}l@{}}Cost per hour (\$)\\ Reliable-Unreliable\end{tabular}} & \multicolumn{1}{l}{$p_i$} \\
\hline
a1.medium  & 2                          & 0.0255                                            & 0.005                                             & 30\%                   \\
a1.large   & 4                          & 0.051                                             & 0.0098                                            & 28\%                   \\
a1.xlarge  & 8                          & 0.102                                             & 0.0197                                            & 25\%                   \\
a1.2xlarge & 16                         & 0.204                                             & 0.0394                                            & 22\%                   \\
a1.4xlarge & 32                         & 0.408                                             & 0.0788                                            & 20\%                   \\ 
\hline
\hline
\end{tabular}
\end{table}

\textbf{\textit{The Interruption Model}}. 
As the demand for unreliable instances can vary significantly over time, the availability of such instances is questioned. An unreliable instance can be interrupted at any time and the allocated capacity is returned to the Cloud provider. Amazon claims that the average interruption probability across all regions and instances is less than 5\%. However, different types of instances are associated with different interruption probabilities \cite{amazonSpot}. In our experimental evaluation, we consider that interruptions may occur at any slot ($S_\tau$) during the execution of tasks in any unreliable resource. 
Table \ref{VMs} depicts the interruption probability for each VM. After an interruption, the corresponding VM does not become available again, unless it is requested from the provider (provisioning and de-provisioning costs are considered). To achieve a fault-tolerant setup, we consider task retries, i.e., revoked tasks along with not running tasks are resubmitted to be scheduled at the same time that the revocation event actually happened.

\textbf{\textit{Performance Metrics}}.
We report on the performance of our model concerning its ability of making correct decisions when deciding the pricing model (reliable or unreliable) for the execution of tasks. We also focus on the workflow's execution time and the overall monetary cost. The performance of the proposed mechanism is evaluated by a set of metrics. We adopt a set of metrics in the following axes: (i) the accuracy of the FLC ($acc$). To measure $acc$, we define the number of correct decisions $\Delta$. To do so, we assume two binary functions $\lambda_1(M_t)$ and $\lambda_2(C_t)$. $\lambda_1$ is equal to unity, if task $t$ is executed to a reliable VM when $HEFT_{ideal}$ is applied, otherwise is equal to zero. Similarly, $\lambda_2$ is equal to unity, if $t$ is executed to an unreliable VM when $GC_{ideal}$ is applied, otherwise is equal to zero. A decision is considered as correct when one of the Eq.(\ref{corr1}), Eq.(\ref{corr2}) holds true. For instance, Eq.(\ref{corr1}) indicates that a decision is correct when $HEFT_{ideal}$ assign $t$ to an reliable VM and, at the same time, the FLC decides to schedule $t$ to reliable VM. The final accuracy of the proposed FLC is measured as follows: $acc=\Delta/|T|*100$.
\begin{equation}
\label{corr1}
\lambda_1(M_t)=1 \quad \&\& \quad FLC_t \to reliable
\end{equation}
\begin{equation}
\label{corr2}
\lambda_2(M_t)=1 \quad \&\& \quad FLC_t \to unreliable
\end{equation}
(ii) the final makespan ($M_{final}$) and the monetary cost ($C_{final}$) generated by the proposed algorithm. Both metrics are normalized using the following equations $normM=M_{final}/M_{lower}$ and $normC=C_{final}/C_{lower}$ respectively. When $normM$ and $normC$ are close to unity the performance of the proposed algorithms is considered as efficient. 

We perform a set of experiments for different $\theta$, $a$ and $b$. $\theta$ varies from 0.1 to 0.9 while for $a$ and $b$ we consider both tight are relaxed upper bounds ranging from 0.5 to 3.0, respectively. 
In total, we conduct 100 iterations for each experiment and report our experimental outcomes for the aforementioned metrics. Experiments where conducted on a Linux server with two 6-core Intel Xeon E5-2630 CPUs running at 2.3GHz.

\subsection{Performance Assessment}
\begin{figure}[h]
\centerline{\includegraphics[width=0.75\linewidth]{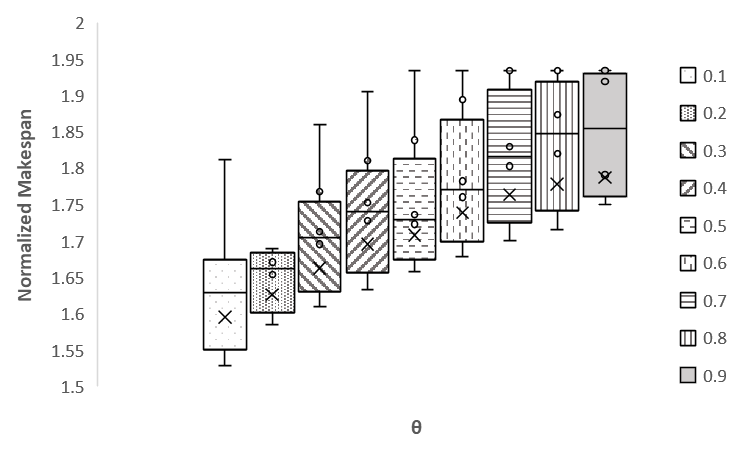}}
\caption{Our evaluation outcomes related to $normM$}
\label{fignM}
\end{figure}

\begin{figure}[h]
\centerline{\includegraphics[width=0.75\linewidth]{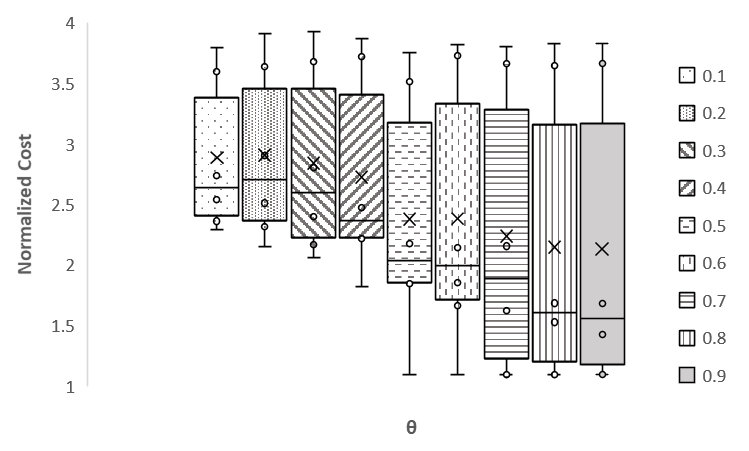}}
\caption{Our evaluation outcomes related to $normC$}
\label{fignC}
\end{figure}

Initially,  we perform a set of simulations for various 
$\theta$ realizations and illustrate its effect on both, $normM$ and $normC$. Recall that when $PMI_i^\tau \geq \theta$, $t_i$ will be scheduled to a reliable resource otherwise the selection of an unreliable resource is the case. In Fig. \ref{fignM} and Fig. \ref{fignC}, we plot the normalized makespan ($normM$) and ($normC$) for different combinations of $a$ and $b$ while $\theta$ varies from 0.1 to 0.9. 
First, it becomes clear that any performance difference is rather small as $\theta$ increases. 
As expected, $\theta$ and $normC$ follow the same trend. 
This is reasonable since as $\theta$ increases the majority of tasks are assigned to unreliable VMs. However, even for the case where $\theta=0.9$, the makespan is less that two times the lower bound. In Fig. \ref{fignC}, $\theta$ is inversely proportional to $normC$. This stems from the fact that high $\theta$ values suggest the use of costly-effective unreliable VMs. 
When $\theta \in [0.5, 0.9]$, we observe cases where the cost is nearly equal to the lower bound. 

\begin{figure}[h]
\centerline{\includegraphics[width=0.75\linewidth]{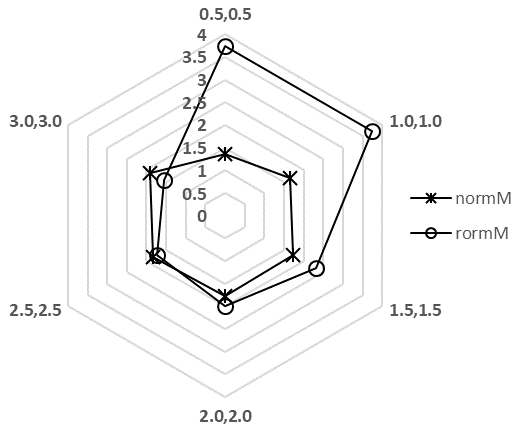}}
\caption{$a \in [0.5-3.0]$ $b \in [0.5-3.0] $}
\label{fig_b1}
\end{figure}

\begin{figure}[h]
\centerline{\includegraphics[width=0.75\linewidth]{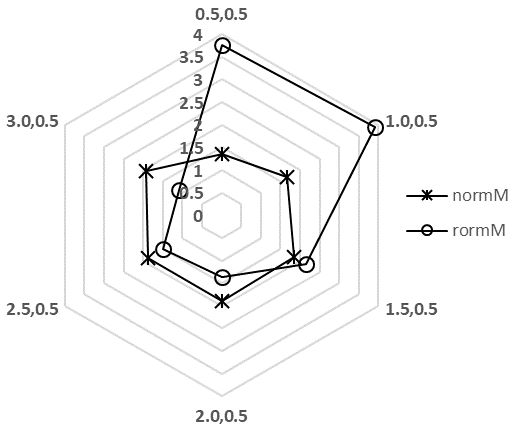}}
\caption{$a \in [0.5-3.0]$ $b=0.5 $}
\label{fig_b2}
\end{figure}

\begin{figure}[h]
\centerline{\includegraphics[width=0.75\linewidth]{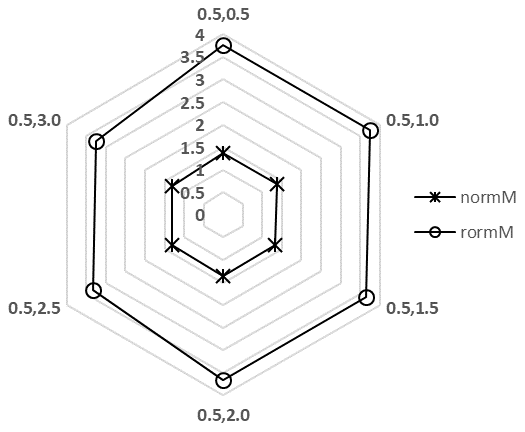}}
\caption{$a=0.5$ $b \in [0.5-3.0] $}
\label{fig_b3}
\end{figure}

In Figs. \ref{fig_b1}-\ref{fig_b3}, we keep $\theta =0.5$ and present the performance of the UDS algorithm
for different combinations of $a$ and $b$ (six in total for each case). We observe that when we expect the performance of the UDS algorithm
to be extremely close to lower bounds ($a=b=0.5$), the proposed algorithm sacrifices cost for execution time. 
This is natural as our FL rule base `suggests' to use reliable VMs when the distance from lower bound is high. On the other hand, the UDS algorithm
favors the cost as the distance from lower bounds increases. 
This is due to the fact that the FLC suggests the use of unreliable VMs when the distance from lower bounds is limited. 
However, the effect on the makespan is relatively minor compared to the benefit on the cost. 
In the experimental scenario where we get $a=3.0$ \& $b=0.5$, we enjoy the best performance related to the cost metric ($normC=1.1$) while the reverse scenario, i.e., $a=0.5, b=3.0$ leads to the best performance for the makespan metric ($normM=1.28$). 
To efficiently perform on both metrics, in parallel, the distance from the upper bound must be moderate (e.g., $a=2.0, b=2.5$). 
\begin{figure}[h]
\centerline{\includegraphics[width=0.75\linewidth]{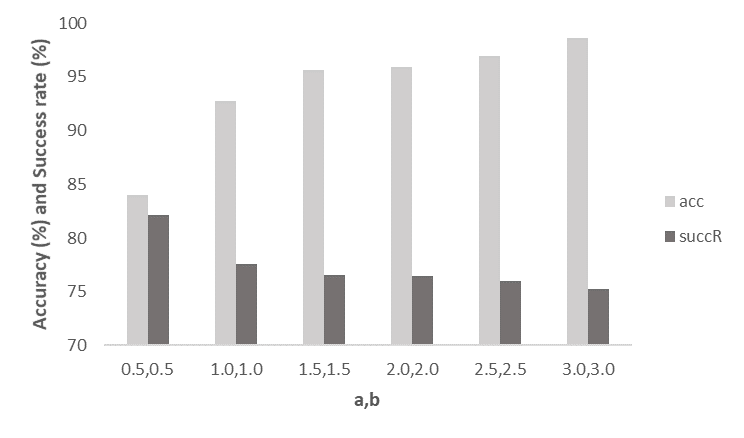}}
\caption{Accuracy (\%) and Success Rate (\%)}
\label{acc}
\end{figure}

Fig. \ref{acc} presents the accuracy ($acc$) of our model as well as the percentage of tasks that have been executed successfully ($succR$). We can see that for high $a$ and $b$, $acc$ is more that 95\%. 
This is also confirmed when we focus on high $a$ and $b$, i.e., in these scenarios, the UDS algorithm
achieves an efficient performance for both metrics, as explained above. However, the number of the successfully executed tasks is decreased as the proposed algorithm 
utilizes more unreliable VMs. In  any  case, the proposed approach is  characterized  by stability as different $a$ and $b$ values have a minor impact in the accuracy of our model.

\section{Conclusions}
In this paper, we tackle the problem of scheduling scientific workflows over distributed heterogeneous resources using different Cloud-based pricing models. To decide on the pricing model, the proposed algorithm (UDS) incorporates a FLC that delivers the  realization of an indicator over which the final decision is made. The discussed indicator shows the efficiency of executing a task to reliable or unreliable resources. Viewing the results in a retrospect, we can argue that UDS tackles both optimization targets i.e., the execution time and the monetary cost achieving a high accuracy (up to 98\%). In the first place of our future agenda is to apply the optimal stopping theory to detect the appropriate time to migrate a task to more-certain resources and semantically cover the heterogeneity of the available resources. 
\label{section:conclusions}

\bibliographystyle{ieeetr}

\bibliography{refs}

\end{document}